\documentclass[9pt,twocolumn,twoside]{opti_preprint}

\setboolean{shortarticle}{false}

\usepackage{soul}

\title{Transmission-matrix-based point-spread-function engineering through a complex medium}

\author[1]{Antoine Boniface}
\author[1]{Mickael Mounaix}
\author[1,2]{Baptiste Blochet}
\author[3]{Rafael Piestun}
\author[1,*]{Sylvain Gigan}

\affil[1]{Laboratoire Kastler Brossel, ENS-PSL Research University, CNRS, UPMC Sorbonne Universit\'{e}s, Coll\`{e}ge de France, 24 rue Lhomond, 75005 Paris, France }
\affil[2]{Ecole Normale Supérieure,  IBENS, CNRS UMR8197, INSERM U1024, 46 rue d’Ulm, Paris,
F-75005 France}
\affil[3]{Dept. of Electrical, Computer, and Energy Engineering, 
University of Colorado, Boulder, Colorado, 80309, USA}

\affil[*]{Corresponding author: sylvain.gigan@lkb.ens.fr}

\begin{abstract}
We report a method to design at will the spatial profile of transmitted coherent light after propagation through a scattering sample.  We compute an operator based on the experimentally measured transmission matrix, obtained by numerically adding an arbitrary mask in the Fourier domain prior to focusing. We demonstrate the strength of the technique through several examples: propagating Bessel beams, thus generating foci smaller than the diffraction limited speckle grain, donut beams, and helical beams.  We characterize the 3D profile of the achieved foci and analyze the fundamental limitations of the technique. Our approach generalizes Fourier optics concepts for random media, and opens in particular interesting perspectives for super-resolution imaging through turbid media.
\end{abstract}

\begin{document}

\maketitle
\pagestyle{plain}
\ifthenelse{\boolean{shortarticle}}{\abscontent}{}

\section{Introduction}
Generating a specific optical point-spread-function (PSF) has been one of the cornerstones of modern microscopy. This is conventionally done by inserting a phase or amplitude mask in the Fourier plane of the imaging system. For instance, Durnin et al. generated Bessel beams using a spatial filter for beam shaping~\cite{durnin1987diffraction}. Nowadays, holographic methods using a spatial light modulator (SLM) are the most versatile~\cite{forbes2016creation,maurer2011spatial}. These techniques allow flexible beam shaping in a wide range of applications such as super-resolution microscopy~\cite{willig2006sted,grover2012super}, 3D microscopy~\cite{planchon2011rapid,Pavani:08}, optical tweezers~\cite{zhang2003optical,Conkey:11,Schonbrun:05} and particle trapping~\cite{dholakia2011shaping}. However, all these studies typically require high quality optics and demand little or no  sample aberrations. 

\begin{figure*} [!ht]
\centering
\includegraphics{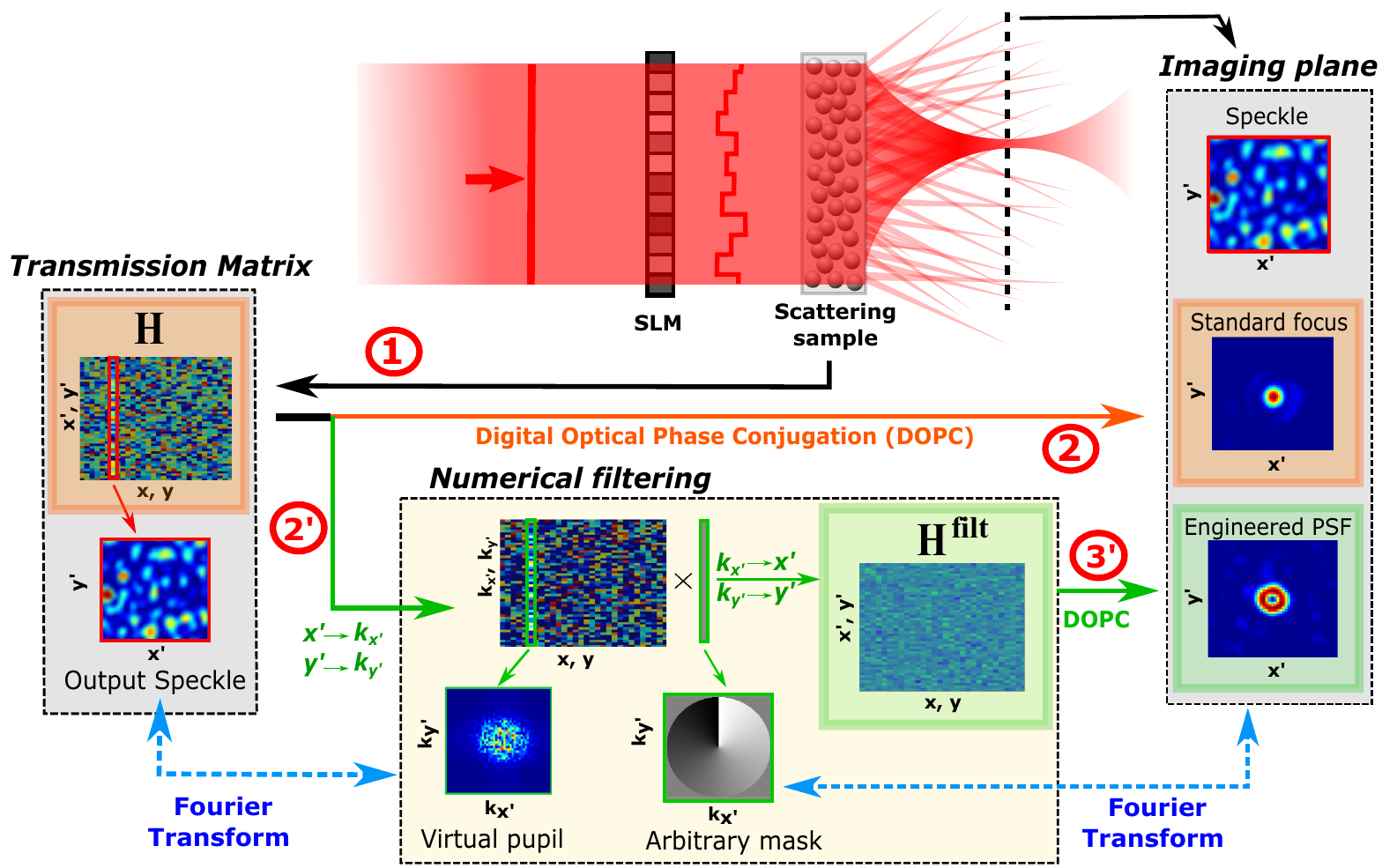}
\caption{General principle for PSF engineering behind a scattering medium. Top: Schematic of the experiment. A Spatial Light Modulator (SLM) is placed on one side of a scattering medium to shape the incident light. The transmission matrix is first measured, then a new operator is calculated. From this operator, we can generate the SLM pattern, enabling focusing with arbitrary PSF after transmission through the sample in the imaging plane. Boxes: Details of each steps to reach arbitrary PSF focusing. \textcircled{\protect\raisebox{-1pt}{1}} The standard Transmission Matrix $H$ characterizes the propagation of light in the scattering medium.\textcircled{\protect\raisebox{-1pt}{2}} The digital optical phase conjugation (DOPC) of $H$ enables focusing in any position of the imaging plane, with a size limited by diffraction~\cite{popoff2010measuring}. \textcircled{\protect\raisebox{-1pt}{2'}} To go beyond the previous approach, a numerical step can be added to control the PSF. With a prior numerical computation of the optical field in a virtual pupil with a Fourier transform operation, the corresponding new operator $H^{\text{filt}}$ is obtained by numerically applying an arbitrary mask onto that pupil.\textcircled{\protect\raisebox{-1pt}{3'}} The DOPC of $H^{\text{filt}}$ allows focusing in the output plane after the medium, the shape of the focus being given by PSF (in this example a donut mode), defined by the Fourier transform of the arbitrary mask. As in~\cite{popoff2010measuring}, the complex focus stands over a speckle background.}
\label{scheme_principle}
\end{figure*}

Light propagation in materials with optical index heterogeneities is affected by scattering. In scattering materials, such as white paint or biological tissue, multiple scattering is at the origin of an intricate interference light field at the output of the medium, also known as speckle pattern~\cite{goodman1976some}. Although the size of a speckle grain is diffraction-limited, this complex interference figure is detrimental for all conventional imaging systems. 
Nonetheless, this scattering process is linear and deterministic and thus even strongly scattering materials can be described by a transmission matrix (TM) ~\cite{popoff2010measuring}. Wavefront shaping techniques have recently emerged as a powerful technique to control the output field, using spatial light modulators or phase-conjugate mirrors~\cite{mosk2012controlling}. SLMs provide up to several million degrees of freedom to design the input field at will, in order to control the propagation of light after the medium. Over the last decade, pioneer works have proven the capability to drastically increase light intensity at one or several output positions of a disordered system such as white paint~\cite{vellekoop2007focusing,popoff2010measuring}, multimode fibers~\cite{papadopoulos2012focusing} or biological samples~\cite{yaqoob2008optical}. However, the resulting focal spots have sizes comparable to a speckle grain, thus diffraction-limited \cite{vellekoop2010exploiting}. Recently, Di Battista et al~\cite{di2015enhanced} overcame this size limit by mechanically inserting an annular mask just after the scattering medium, prior to iteratively optimizing the focus intensity at a distance via wavefront shaping. Due to the filtering of the low spatial frequencies, the resulting speckle and optimized spot was narrower than the initial speckle. After removing the filter, the narrow spot - effectively a Bessel-like beam - remained intense, over a background speckle pattern wider than the focus. 

Herein we report the first formulation of a TM-based operator that enables deterministic focusing after propagation through the medium, with a controllable PSF. We build this new operator by numerically applying a well-chosen mask (that can be arbitrarily designed in amplitude and/or phase) in a virtual Fourier plane of the output modes of the experimentally measured transmission matrix. We then demonstrate experimentally that a focus with the corresponding PSF can be obtained after the medium, by performing digital phase conjugation on this operator. To demonstrate the robustness and wide applicability of our technique, we generate and characterize a variety of useful PSF. Firstly, we generate a Bessel beam focus using an amplitude annulus mask, and show that its  central FWHM is narrower than the size of a speckle grain, therefore demonstrating deterministic sub-speckle focusing without mechanical masking as in~\cite{di2015enhanced}. We also demonstrate donut-mode generation with various topological charges, and helical foci. Characterization of the axial properties of Bessel and helical PSFs shows he potential of the technique for 3D wave control.

\section{Principle of the experiment}
A particular class of wavefront shaping methods relies on the measurement of the optical transmission matrix (TM), denoted $H$ in this paper, which contains the relationship between the input field and the output field~\cite{popoff2011controlling}. Its complex coefficients $h_{\textbf{X'X}}$ connect the optical complex field at the output [$\textbf{X'}=(x',y')$ CCD pixel coordinates] to the input field [$\textbf{X}=(x,y)$ SLM pixel coordinates] by:
\begin{equation}
E_{\textbf{X'}}^{\text{out}}= \sum_{\textbf{X}}h_{\textbf{X'X}}E_{\textbf{X}}^{\text{in}}
\end{equation}
Experimentally the TM is measured by displaying a set of input fields on the SLM and recording the corresponding output fields on the CCD, and requires the medium to be stable during the whole measurement process (a few minutes in our case).
The digital phase conjugation of the TM, $H^{\dagger}$ where $\dagger$ stands for the conjugate transpose, enables focusing at any output position~\cite{popoff2010measuring}, or scanning the focus~\cite{caravaca2014high}. The resulting spot has a size given approximately by the spatial correlation of the output speckle, i.e. diffraction-limited~\cite{goodman1976some,vellekoop2010exploiting}. It effectively sets to a common phase all contributions arriving at this position. However, it is possible to completely tune the phase and amplitude distribution of the k-vectors forming this focus, and therefore control at will the PSF.
In Fig.~\ref{scheme_principle} we detail how we build a new operator based on the experimentally measured TM to generate an arbitrary mask, and generate the corresponding PSF at the focus.
We first numerically perform  a two-dimensional spatial Fourier transform on the TM, noted $\mathfrak{F}_{2D}$, of every output field. We define $\hat{h}_{\textbf{K'X}}=\mathfrak{F}_{2D}(h_{\textbf{X'X}})$, where ${\textbf{K'}}=(k_{x'},k_{y'})$ is the wave vector associated to $\textbf{X'}$. This numerical operation is equivalent to compute the TM in a Fourier plane of the output imaging plane.
In order to generate a given PSF, we then multiply in the k-space the field in the pupil plane by  mask $M$ (amplitude and/or phase), corresponding to this PSF. 
We thus obtain a new numerically filtered coefficient in the Fourier domain:
\begin{equation} \label{filtering}
 \hat{h}_{\textbf{K'X}}^{\text{filt}}=\hat{h}_{\textbf{K'X}}\times M(k_{x'},k_{y'})
\end{equation}
We then return to the spatial domain $\textbf{X'X}$ by taking the inverse Fourier transform of $\hat{h}_{\textbf{\textbf{K'X}}}^{\text{filt}}$:
\begin{equation}
 {h}_{\textbf{X'X}}^{\text{filt}}=\mathfrak{F}_{2D}^{-1}( \hat{h}_{\textbf{K'X}}^{\text{filt}})
 \end{equation}
The resulting operator ${H}^{\text{filt}}$ can now be used instead  of ${H}$  to perform focusing and scanning of a focus, but with the chosen PSF. 

\begin{figure}[t]
\centering
{\includegraphics[width=\linewidth]{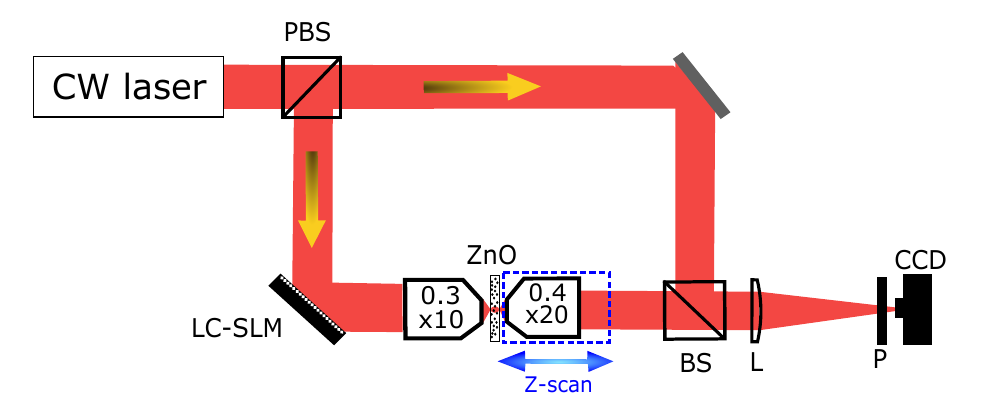}}
\caption{Experimental setup. A cw laser ($\lambda=$ 800nm) illuminates an SLM that modulates the wavefront of light before propagation through a multiple scattering medium (ZnO nanoparticles). Transmitted light is collected with another microscope objective, placed onto a motorized stage to scan the z-axis. The output speckle pattern is recombined with a reference plane wave from the same laser, with a beam splitter (BS). One polarization state of the output field is selected with a polarizer (P) and imaged on a CCD camera. (L): lens (focal distance 200mm)}
\label{setup}
\end{figure}

\section{Experimental demonstration}
Fig.~\ref{setup} sketches the experimental setup to measure the transmission matrix ${H}$, compute ${H}^{\text{filt}}$, and generate and characterize the formed PSF in three dimensions. A cw laser ($\lambda=800$ nm, MaiTai, Spectra Physics) is split between a reference path and a sample path. In the sample path, a phase-only SLM (LCOS-SLM, Hamamatsu X10468) subdivided in $64\times64$ macropixels is conjugated with the back focal plane of a microscope objective, which illuminates a scattering medium made of ZnO nanoparticles (thickness $\simeq$ 100µm).  another microscope objective is used to image the transmitted speckle . The objective is placed onto a motorized stage (Thorlabs, Z825B) to scan the imaging plane axially. For the TM measurement, the output beam is recombined with the reference on a beam splitter and the hologram is recorded on a charged couple device (CCD) camera (Allied Vision, Manta G-046).The reference beam is blocked during focusing and PSF characterization.

\begin{figure}[!t]
\centering
{\includegraphics[width=\linewidth]{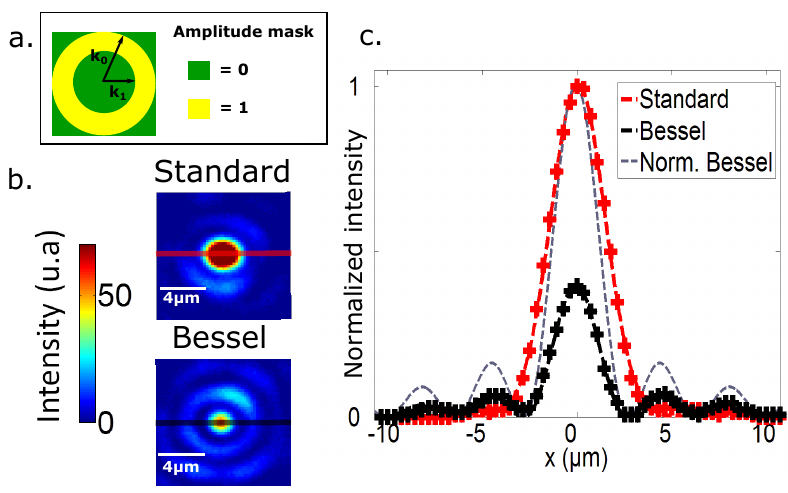}}
\caption{Sub-diffraction focusing with the propagation of Bessel-like beam - Experiment - (a) The mask applied during computation of the effective focusing operator is an amplitude annulus, with inner radius $k_1= 0.58$ $k_0$, with $k_0$ the pupil-width in k-space. (b) Comparison between standard (with $H$) and Bessel-like (with ${H}^{\text{filt}}$ obtained using mask (a) ).  Although standard focusing is brighter, the Bessel-like has a narrower FWHM. (c) Intensity profile of both standard and Bessel-like focusing. Bessel-like focusing is narrower by $23\%$ in FWHM.}
\label{bessel}
\end{figure}

We first demonstrate the generation of a Bessel-like beam using an annular mask. The irradiance profile of the ideal beam is described by a zero-order Bessel function of the first kind, which propagates with an associated complex field $\propto{J}_{0}(k_rr)\exp{(-ik_z z)}$. The experimental result is represented in Fig.~\ref{bessel}. The applied mask during the numerical filtering step is an annular amplitude mask, with an inner ring size $58\%$ of the pupil size. This value has been found to be a good compromise between the loss in intensity at the output, the ability to detect the foci over the background speckle, and a significant narrowing of the FWHM of the central spot, as with a mechanical mask~\cite{di2015enhanced}. Using $\big({H}^{\text{filt}}\big)^{\dagger}$, we obtain in the imaging plane a Bessel-like focus, standing over a background speckle. We compare its FWHM to a standard focus obtain with the standard TM $H^{\dagger}$. In the same experimental conditions, the Bessel-like central lobe FWHM is $23\%$ narrower than a standard speckle focus grain. Both intensity profiles are  fitted to an ideal Bessel and Gaussian profile, respectively.
The central lobe of a Bessel beam is surrounded by a decaying set of side-lobe rings. Each lobe carries approximately the same amount of energy as the central spot, whereas Gaussian beams contain 50\% of their total energy within their FWHM~\cite{Durnin:88}. Achieving a smaller FWHM than a diffraction-limited focus entails penalties such as loss of energy in the central peak (40\% lower than a standard beam). Inherent to wavefront shaping in complex media techniques, the PSF is not perfect, but rather stands over a speckle pattern that remains in the background, with the peak to background ratio proportional to the number of input modes~\cite{vellekoop2007focusing}. 
In the generation of engineered PSFs, an extra penalty is added due to the low transmission of the virtual annular mask, which steeply decreases the energy in the focus while the background remains the same. In order to detect the focus with sufficient SNR, a high number of degrees of freedom, i.e. SLM pixels, is required, here we used $N=4096$ input pixels. We therefore demonstrate sub-speckle focusing after propagation through a scattering medium, using a Bessel-like beam.

\begin{figure}[t]
\centering
{\includegraphics[width=\linewidth]{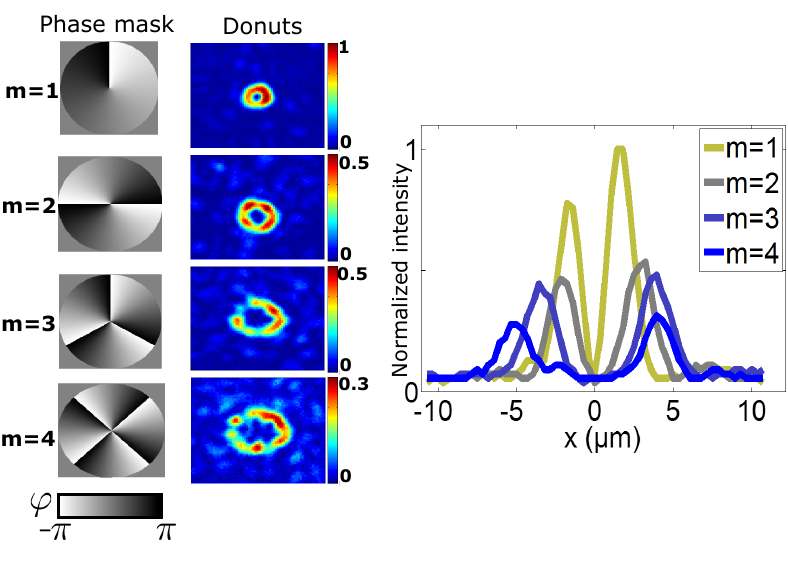}}
\caption{Generation of donut modes of various diameters after propagation through the scattering medium. Diameter increases with topological charge $m$ from 1 to 4 (related to the corresponding  Laguerre-Gauss beam $LG_0^m$).   Applied masks in the pupil plane (left), intensity distribution retrieved with a CCD camera (center), and related intensity profile for each order (right).}
\label{donuts}
\end{figure}

Since any arbitrary phase and/or amplitude mask can be computed and placed onto the virtual pupil field, we also demonstrate the generation of donut-shaped beams, which are closely related to single-ringed Laguerre-Gaussian beams ($LG_0^m$). The corresponding mask to be applied in the numerical filtering step is a spiral phase plate distribution presenting a continuous and gradual phase change from $0$ to $2m\pi$ around the optical axis, where $m$  is to the integer number of $2\pi$ cycles in the pupil plane~\cite{ganic2003focusing} and corresponds to a topological charge.
The ring pattern dimensions increase with $m$. Experimental results of focusing ($LG_0^m$) beams, with $m$ from 1 to 4, are presented in Fig.~\ref{donuts}, using $N=1024$ SLM pixels. The transmittance M  of the numerical phase mask in the Fourier domain applied during the numerical filtering step, as defined in Eq.~\ref{filtering} and illustrated in Fig.~\ref{donuts}, reads : 

\begin{equation}
M^{\text{Donut}}(k,\theta)= \text{circ}\Big(\frac{k}{k_0}\Big) \exp{(i m \theta)}  
\end{equation}
where $k_0$ is the pupil size in k-space.
Just as for with Bessel-like beam of Fig.~\ref{bessel}, the focus stands over a background speckle pattern. When $m$ increases, the intensity is distributed over a wider area in the imaging plane. Therefore, for a fixed number of SLM pixels used, the total energy in the targeted area is about the same but distributed over a bigger area, while the background speckle remains at the same average intensity.

\begin{figure}[t]
\centering
{\includegraphics[width=\linewidth]{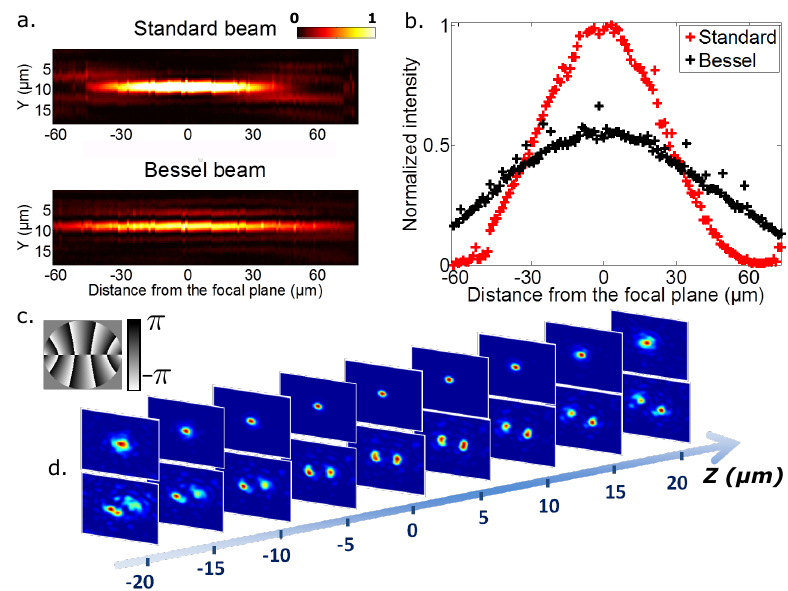}}
\caption{z-axis properties of 3D beams. (a) The Bessel-like beam diverges more slowly than the gaussian depth of focus, related to the Rayleigh range ($z_r$) of the microscope objective. (b) Corresponding intensity profiles along z-axis of both standard and Bessel-like foci. (c) Phase mask applied in the pupil plane for double helix (DH) focusing. (d) Corresponding double helix beam along z-axis. $z=0$ corresponds to the focal plane axial position.}
\label{zaxis}
\end{figure}

Several beam profiles, such as Bessel beams have also very interesting propagation properties along the Z-axis. Our setup enables the scan of the Z-axis thanks to a motorized stage under the collecting microscope objective. Experimentally, we can scan over a few Rayleigh ranges ($z_R = 10 \mu m$) on both sides of the focal plane. 
In Fig.~\ref{zaxis}a , yOz cross sections of  Bessel-like beam obtained in the same conditions as in Fig.~\ref{bessel}a are reported. 
As expected, we prove without ambiguity that the generated Bessel-like beam has a longer depth-of-focus during propagation along the z-axis relative to standard Gaussian beams, for which the depth of focus is the Rayleigh length. Here, we observe experimentally that our Bessel-like beam has a depth of focus 1.7 times longer than the Gaussian focus. 

As a last example, we demonstrate the generation of double-helix point spread functions (DH-PSF). This 3D-design has 2 dominant lobes in the imaging plane, whose angular orientation rotates with the axial ($z$) position~\cite{grover2012super,Pavani:08}.
This profile is realized by applying a particular phase mask during the numerical filtering step leading to $H^{\text{filt}}$, illustrated in Fig.~\ref{zaxis}c, which reads~\cite{grover2012super}:

\begin{equation}
M^{\text{DH}}(k,\theta)= \text{circ}\Big(\frac{k}{k_0}\Big) \exp{\Bigg[i  \text{arg}\Big(\prod_{j=-M}^M(k\exp{(i\theta)} - k_j\exp{(i\theta_j)})\Big)\Bigg] }
\end{equation}
where M=($N_v$-1)/2, $N_v=9$ is the number of vortices, $(k_j,\theta_j)$ the position of the $j$th vortex and $k_0$ the pupil size. A stack of images taken on both sides of the focal plane illustrates the rotation of the DH-PSF in  Fig.~\ref{zaxis}d.
The pattern rotates approximately linearly with $z$, with a negative angle for negative $z$. The reference $z=0$ corresponds to the focal plane position. The DH rotates from -30$^{\circ}$  to +30$^{\circ}$ between -2$z_R$ and +2$z_R$. This type of profiled-PSF has been used for 3D localization with a single shot, where the rotation of the imaged spot is related to its depth~\cite{pavani2009three}.

\section{Discussion}
We have demonstrated focusing with various PSFs, defined by arbitrary controlling a mask in phase and amplitude in the virtual Fourier domain. It is to be noted that, as in all methods relying on focusing through a complex medium, the focus stands over a background speckle resulting from the incomplete phase conjugation \cite{vellekoop2007focusing}, the signal to background of the focus increasing linearly with the number of segments controlled on the SLM. However, our technique goes well-beyond spatial-shaping techniques relying on intensity optimization in the imaging plane;  namely, our approach allows for fine control of the focus shape, in amplitude and phase, that cannot simply be achieved using optimization approaches on the intensity. In the latter, the maximal intensity scales inversely with the number of target points \cite{vellekoop2007focusing} or the size of the focusing area, as in acousto-optic and photoacoustic techniques \cite{conkey2015super,wang2012deep}. Similarly, in our approach, the maximal intensity in the PSF decreases with its complexity, but the amount of energy in the PSF remain the same. 

This  PSF engineering method through complex medium is also not limited to amplitude and phase modulation in the Fourier domain, and could be for instance extended to arbitrary polarization mask if measuring a polarization-resolved transmission matrix~\cite{Tripathi:12}, and even to more complex spectrally-dependent PSFs thanks to the spectral dependence of the speckle~\cite{PhysRevLett.116.253901,conkey2012color}. One can therefore generate PSFs that could not easily be realized with physical masks or SLMs. Moreover, the Fourier domain of the plane of interest may not be accessible in practice, as in~\cite{di2015enhanced} where the amplitude mask was placed after the medium and the speckle observed at a distance. In contrast, our technique would work in any plane, even at the output surface of the medium. 
Furthermore, a single transmission matrix measurement can be used to focus at different positions, and allows rapid switching between various PSFs, as the numerical filtering step and phase conjugation are realized a posteriori. Using a simple quadratic phase mask, the focusing plane can also be translated longitudinally at will. 

On the downside, in order to perform an accurate spatial Fourier-transform, the reference beam during the transmission matrix measurement needs to be a well-defined plane wave, which requires a reference arm with interferometric stability during the measurement process. Additionally,  the resolution of the generated mask is  related to the sampling in the Fourier domain, and therefore depends mostly on the sampling in the imaging plane (on the CCD): generating an accurate mask requires sampling at least a few CCD pixels per speckle grain over an extended output spatial region of at least a few times the PSF size. Note that, however, the resolution of the mask is advantageously independent of the number and resolution of the SLM segments, that only affect the signal to background of the focus after phase-conjugation. It is also worth pointing out that the signal to background is crucially affected by the total transmission of the virtual mask (as pointed out by~\cite{di2015enhanced} for physical masks).

\section{Conclusion}
In conclusion, we have reported the first formulation of an operator, built upon the experimental transmission matrix, that enables deterministic focusing of any arbitrary PSF after propagation through a scattering sample. We have illustrated the strength of this technique by generating Bessel, "donut" and Double Helix beams through a scattering sample with a simple use of this new operator, and characterizing their transverse and longitudinal properties. The method can readily be extended to other complex media, from biological tissues to multimode fibers. The possibility of arbitrarily generate complex PSF through or in scattering media opens up new perspectives for many fields, in particular for 3D and super-resolution microscopy and for optical manipulation and trapping~\cite{vcivzmar2010situ}.

\paragraph{Funding Information}
This research has been funded by the European Research Council ERC COMEDIA (278025). S.G. is a member of the Institut Universitaire de France. R.P. acknowledges NSF support under awards 1611513 and 1310487.

\paragraph{Acknowledgment}
The authors thank Andr\'{e}ane Bourges, Hugo Defienne and Gaetan Gauthier for discussions. 

\bibliographystyle{apsrev4-1}
\bibliography{biblio_psf_engineering}

\end{document}